# Making Videos Accessible for Blind and Low Vision Users Using a Multimodal Agent Video Player

Adriana Olmos*

GOOGLE, MONTREAL, QUEBEC, CANADA

Anoop K. Sinha*

GOOGLE, Paradigms of Intelligence, MOUNTAIN VIEW, CALIFORNIA, UNITED STATES

Renelito Delos Santos

GOOGLE RESEARCH, MOUNTAIN VIEW, CALIFORNIA, UNITED STATES

Ruben Rodriguez Rodriguez

GOOGLE, MOUNTAIN VIEW, CALIFORNIA, UNITED STATES

James A. Landay

GOOGLE, Paradigms of Intelligence, MOUNTAIN VIEW, CALIFORNIA, UNITED STATES
COMPUTER SCIENCE DEPARTMENT, STANFORD UNIVERSITY, STANFORD, CALIFORNIA, UNITED STATES

Sam S. Sepah

GOOGLE RESEARCH, MOUNTAIN VIEW, CALIFORNIA, UNITED STATES

Philip Nelson

GOOGLE RESEARCH, MOUNTAIN VIEW, CALIFORNIA, UNITED STATES

Shaun K. Kane

GOOGLE RESEARCH, BOULDER, COLORADO, UNITED STATES

**ABSTRACT**

Video content remains largely inaccessible to blind and low-vision (BLV) users. To address this, we introduce a prototype that leverages a **multimodal agent** — powered by a novel conversational architecture using a multimodal large language model (MLLM) — to provide BLV users with an interactive, accessible video experience. This Multimodal Agent Video

Player (MAVP) demonstrates that an interactive accessibility mode can be added to a video through multilayered prompt orchestration. We describe a user-centered design process involving 18 sessions with BLV users that showed that BLV users do not just want accessibility features, but desire independence and personal agency over the viewing experience. We conducted a qualitative study with an additional 8 BLV participants; in this, we saw that the MAVP's conversational dialogue offers BLV users a sense of personal agency, fostering collaboration and trust. Even in the case of hallucinations, it is meta-conversational dialogues about AI's limitations that can repair trust.

# 1 INTRODUCTION

Video is a pervasive medium for information, learning, entertainment, and communication. However, for millions of blind and low-vision (BLV) individuals, video presents a significant accessibility barrier [42]. One increasingly common accessibility feature, Audio Description (AD) [55,71], involves a narrator describing visual elements during pauses in the dialogue. While beneficial, standard AD has critical limitations [37,43,58], including that users can only access what was included in the pre-recorded audio track, leaving them unable to ask for clarification, deeper context, or more detail about a specific scene or object that piques their interest.

Static, limited AD can lead to an incomplete and disempowering video experience for BLV users [37,58], and many BLV users avoid video content altogether, per our user interviews. This limitation led to our core research question: how can we augment video in a way that shifts the experience from passive consumption to a user-led dialogue? We also hypothesized that the recent introduction of multimodal large language models (MLLM) would allow us to go beyond past research explorations of interactive video exploration and create a more comprehensive and novel experience using MLLMs.

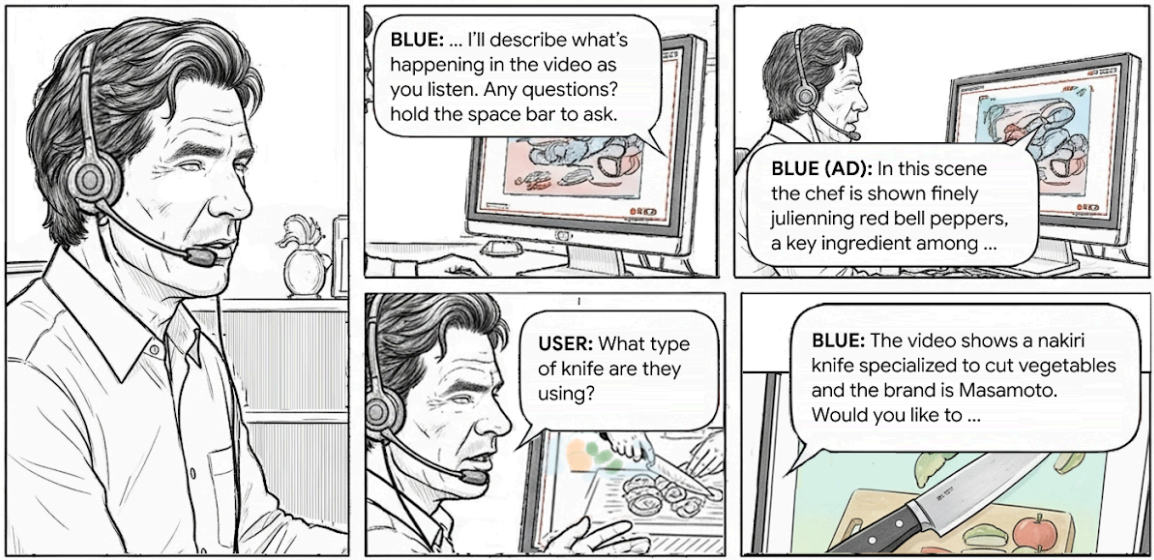

Figure 1. Conceptual overview of the Multimodal Agent Video Player (MAVP). The agent, named "Blue," helps a blind or low-vision user engage in a conversational dialogue about video content by reading aloud audio descriptions and answering questions, providing detailed, on-demand information about its visual elements.

Thus, we introduce the Multimodal Agent Video Player (MAVP) system with a conversational dialogue interface designed for BLV users (Figure 1). Enabled by recent advancements in MLLMs, the system changes the experience from passive listening to interactive dialogue through natural language input and audio output. The interface allows users to:

1. **Adapt the level of detail** in the automated audio descriptions (e.g., "Give me descriptions with more detail.")
2. **Pause the video and ask specific questions about the video** (e.g., "What is the character wearing?" or "Describe the room in more detail." or "What's the name of the main character again?")
3. **Ask specific questions** of the broader context **beyond the video** (e.g., "What is the average price of a speaker like the one on the video?" or "Where can I purchase a painting like that one in the background?")
4. **Control the video playback and configuration experience** entirely through voice commands (e.g., "Go to the part in the video where the character changed to a black dress." or "Switch to dark mode.")
5. **Receive interactive guidance** to better understand the application's functionality (e.g., "What other controls are available?" or "What else can I do?")

The MAVP was developed in an iterative design process with 18 design feedback sessions. The MAVP was evaluated in a qualitative user study with 8 total participants (seven BLV participants and one sighted participant).

This paper contributes:

- Case study of the design and development journey for building an accessible video player for the BLV community using a multimodal agent system via MLLMs
- Insights from the study about the relationship between accessibility and autonomy, showing that video accessibility emerges not from audio descriptions or interactive features alone, but from a conversational dialogue between the user and the MAVP system. This conversational dialogue can give BLV users a sense of personal agency, fostering collaboration and trust.
- Insights about how users react to AI output, including hallucinations, and how meta-conversational dialogues with AI about AI's limitations can repair trust.

While many of the MAVP's features have been implemented in isolation (see Section 2), this work is noteworthy for combining state of the art multimodal AI with conversational interaction and control. The MAVP serves as a proof point that conversational interfaces using multimodal agents are an emergent pillar for the design of next-generation accessibility — in which accessibility is more than "accessing features" and is instead an interactive, empowering experience.

## 2  RELATED WORK

Our work on the MAVP is informed by 1) prior research in video accessibility, 2) research in interactive systems for video for BLV users, 3) research in multimodal agents for video exploration.

### 2.1 Video Accessibility

Audio descriptions are a cornerstone of video accessibility, providing narrated explanations of visual content to ensure that BLV users can comprehend media [37,42]. Prior research has found that BLV users comprehend videos with audio descriptions differently than unannotated videos [18,27]; the level of comprehension is affected by factors such as the level of detail in visual scenes, the delivery style of the narration, its emotional impact, and other factors [30]. Particularly with low-quality audio descriptions, users can experience mental fatigue or confusion [43].

Guidelines for AD include making narrations that are objective and concise, illuminating key visual information without overlapping with dialogue or other critical audio cues [11,20,55]. However, the high cost and manual effort required to produce professional-quality AD has limited its availability, leaving a vast amount of video content inaccessible [71]. To address this limited availability, researchers have explored various approaches to easing the authoring of AD:

- **Manual and collaborative authoring:** Tools like Rescribe [47] and YouDescribe [50,72] were developed to simplify the creation of AD by providing user-friendly interfaces. Other work has focused on collaborative and community-driven approaches to authoring descriptions for various content, from user-generated videos to livestreams [9,33,54,63]. Recently, CoSight has explored crowd input to collect additional descriptive comments for videos [61]. While these tools are valuable, they still rely on manual effort that is difficult to scale, even with tools to support feedback for the manual process [40].
- **Automated and semi-automated generation:** Advances in AI have led to automated or semi-automated systems for generating AD for a range of videos, including traditional films and user-generated content [7,12,22,63]. These systems can accelerate the creation process and provide usable audio descriptions, but can lack the contextual richness and nuance of human-authored descriptions, a limitation that has been seen in images [57] and automatically generated alt-text for images [65]. However, as ML models have improved, the quality of AD has been improving over time and rapidly recently due to MLLMs [23]. For the related process of non-visual video editing for BLV users, AVScript [24] has explored AI model-generated descriptive transcripts that combine captions and visual descriptions into a single document, providing an "audio-visual script."
- **Extended and alternative descriptions:** For content with continuous audio and no natural pauses, the Web Accessibility Initiative [11,60] suggests using extended AD, which pauses the video to make time for longer descriptions. Prior work has explored giving users control over the playback of these extended descriptions; for example, Rescribe allowed users to choose between inline, extended, or hybrid descriptions [47]. Describe Now has explored allowing user driven customizations of the level of detail of descriptions to enhance the AD experience for BLV users [13].

Our work directly builds on these approaches, with the MAVP using a MLLM to automatically generate usable descriptions, showing a pathway using a MLLM prompting that has potential to scale availability of AD. Innovatively, our use of the MLLM demonstrates automatic AD description that includes the ability to interpret scenes across multiple frames, which is helpful for insights into video content, and a way to use the MLLM to create a dense index of the entire video for high accuracy, interactive dialogue, with a retrieval augmented generation (RAG) approach [35].

## 2.2 Interactive Video Systems for BLV users

Beyond audio descriptions, interactive exploration, where users are able to ask questions about scenes, has shown potential to improve accessibility for visual content, such as in the SPICA project [43]. SPICA had limits in the objects that could be identified in scenes and depth of descriptions by the ML models available at the time of that research. Our work shows that a modern MLLM can now enable a more fluid interactive dialogue and novel experience across an entire video.

Further, hierarchical summaries have been used to help BLV users browse and skim long texts, audio recordings, and videos [58], providing a high-level overview while allowing users to drill down into details on demand. As an example, SceneSkim provides video summaries at multiple levels of detail for film professionals [45], and Video Digests offers chapter summaries for those watching educational videos [46]. Recognizing that BLV users often do not want an exhaustive description for every image, systems like ImageExplorer allow BLV users to selectively navigate image details through multi-layered touch exploration [34]. Similarly, Slidecho supports non-visual exploration of presentation videos by extracting slide content and aligning it with the speaker's narration [48]. Additionally, there have been explorations of supporting BLV users exploring 360-degree video [31]. There have also been earlier explorations with human-in-the-loop for video, where humans can edit generated descriptions [66].

Our work targets a more fully interactive paradigm for supporting BLV users' video watching. This has challenges that we addressed. Unlike static images or slides, video content is temporal and requires optimizing for both comprehension and user engagement across the current scene and the entire video. Other systems have explored real-world exploration across time with tools like Seeing AI [67], or with humans-in-the-loop tools such as Be My Eyes [69] and Aira [68]. Early work on the use of a MLLM to provide interactive real-world scene descriptions for blind and low-vision (BLV) users in a visual interpretation application [25,26] has shown both potential value and trust challenges. Recently, Vid2Coach used a MLLM to help guide a BLV user going through a how-to cooking task [28]. Likewise, StreetReaderAI used an MLLM to help a BLV user interactively explore online maps [19]. Vid2Coach and StreetReaderAI have analogous usage of MLLMs as our work, though in different application domains. With these in mind, our work demonstrates allowing BLV users to engage in dialogue on the entire content of videos using AI, from scene details to the entire video to information beyond the video — new interaction enabled by recent MLLMs.

## 2.3 Multimodal Agents for Video Exploration

Efforts on foundation models supporting video understanding have a long thread of research that intensified around when Visual Language Models (VLMs) came into focus with CLIP [52] and Flamingo [1]. These efforts have over the last few years been productionized in many major MLLMs [24,44]. Numerous efforts in interactive Q&A for videos have shown increased performance over time [16,36,38,62]. Our work builds on these foundations, utilizing a production MLLM for building a multimodal agent for video exploration to support BLV users.

Earlier work on interactive agents for exploring video content for BLV users had accuracy limits largely due to the limits of earlier AI models, such as in the AI versions of the NarrationBot+InfoBot system [56]. More recent efforts have shown that AI can be effective in interactively asking questions on specific frames, as in SPICA [43], and generating multiple levels of detail for frame descriptions, as in Describe Now [13]. Describe

Now's multiple levels of details for on-demand AD for scenes and used an early MLLM for generation. The MAVP goes further to support conversational interaction and dialogue in a way that changes user interaction and engagement.

The MAVP system builds on these works and shows a recent MLLM that can support creating usable automatic audio descriptions <u>and</u> can support interactive Q&A across an entire video, creating a more complete, novel user experience. As we saw in our study, the combination provides a user-friendly, supportive, more seamless experience that can empower BLV users with a greater sense of personal agency, going beyond basic accessibility to the video content.

## 3 USER-CENTERED DESIGN PROCESS

We developed the initial design of the MAVP through 18 participant feedback sessions with 15 different BLV individuals. Even though the participants could not see the prototype, they served as co-designers through a collaborative moderation process led by the designers [4,14], including participatory dialogs [53] and voice interface co-design [10,53]. Moderators were trained to use inclusive language and adapt to the diverse needs of participants and create a supportive and respectful environment.

This process supported the research team in understanding how to iterate on the prompts to design the agent and adapt the MLLM to address the spectrum of needs of BLV users. Out of the box, the baseline behavior of the MLLM was insufficient for the needs of BLV users, as MLLM production models that we tried appear trained with the assumption that users are sighted. Careful prompt engineering improved usability significantly and rapidly, meeting the needs of BLV users with usable interaction by the end of this design phase.

### 3.1 Formative research with BLV Users

#### 3.1.1 Needs assessment and ideation interviews

The needs assessment phase started with a series of foundational interviews with 10 participants from the BLV community with a spectrum of levels of vision. The conversation focused on their most recent video-watching experiences across online video platforms (e.g., YouTube, Netflix, etc.), probing for details on their use of existing tools like audio descriptions. Each interview concluded with the research team presenting new concepts around interacting with video content using low fidelity visuals and audio snippets and ended with an open-ended discussion and mini-brainstorm about their ideal future video experience. The protocol was designed to explore current video consumption behaviors and contemporary pain points, find any updated insights from past research studies of BLV video pain points [30,37], and come up with design ideas to address them.

*Takeaway: The most important insight from our formative interviews was that users wanted a system that would increase their independence and personal agency in watching videos.*

Participants told us about needing to request help from others for watching videos, of the challenges of accessing features and of navigating web sites, and of not being able to understand the videos that they were "watching" (more accurately, "listening to," for totally blind users). While some of these issues have been raised in prior research, they have largely not been addressed yet in contemporary video platforms.

Some specific barriers raised by participants were:

- **Reliance on audio:** For example, a participant mentioned, "*I don't recognize people's faces and I don't recognize people's voices*" which makes it difficult to follow a timeline if they don't know who is doing what. Another participant said, "*If there's no AD, I miss a huge part of the story.*" Supplementing what was covered in past research studies, we heard in conversations that these contributed to feelings of disempowerment.
- **Content-specific needs:** Participants expressed that the required type of AD and playback controls varied depending on the video content (e.g., DIY videos, movies, news, music videos) as seen in past studies [28,39,49], and that they wanted a way to maintain a continuous understanding of what was happening in the video. One participant described this frustration with a DIY video saying to cut vegetables "like *this*," which forced them to *"guess or ask someone else to find out"* what was meant by "*this*".
- **AD voice quality matters:** Participants showed a greater tolerance for artificial voices in online videos (e.g., DIY content) compared to movies. For movies, high-quality AD had two key characteristics: a natural narrator's voice and a tone that matched the movie's mood. Supplementing was covered in past studies [47], some even described turning off AD entirely when the voice sounded too robotic.

There were additional supplemental insights that came up in our formative interviews that we found important for our design:

- **Difficulty browsing video settings:** Past research has talked about the difficulty of navigating in videos [29]. Additionally, the effort required to navigate video site accessibility settings, requiring finding accessibility settings or navigating with a screen reader, browsing, and finding additional information often caused participants to give up on watching videos altogether. One participant shared this frustration, stating, "*If I'm not provided a [video] link, then I usually don't go searching for it...*"
- **Spectrum of multimodal needs:** The needs of participants with low vision (LV) varied greatly given varying visual profiles, and their use of tools like magnifiers or audio descriptions depended on the specific content and context of the video. As one participant with low vision noted, they use a combination of regular audio, AD, and captions on and off because *"each one tells me something"* and they need all three modes at different times or combined to get the information they need.

The frustrations from video watching experiences were high to the point where many commented it was disempowering. A user commented: *"Honestly, video is so rough for me... I don't do it very often just because of the amount of work involved and I don't have someone that can actually help me do it."* Despite their significant frustrations, participants showed a strong interest in online videos to be able to follow this important media type. Our main takeaway was that personal agency was a fundamental need and a design target.

As we saw in the low fidelity explorations and brainstorms with users in these interviews, we hypothesized that reaching that target would be possible with a conversational AI system, provided it was one that was comprehensive, fluid and user-friendly in supporting users. Beyond the individual features in our low fidelity explorations, natural language understanding closer to the level of help a BLV user might get from a human when watching videos could be the enabler of greater independence. Thus, a comprehensive system would be needed to approach that design goal, and the system would include several key features, based on consistent suggestions from across the participant group during those explorations:

- **Adaptive AD:** Audio descriptions should adapt based on the video's content type and the participant's preferences.
- **On-demand questions:** The ability to ask questions on demand about the video's content would increase understanding and engagement.
- **Intuitive access to controls:** Participants need quick and intuitive access to video playback controls via voice.
- **Multimodal experiences:** LV participants expressed a preference for combined visual and auditory experiences, so they would not have to rely on a single modality to receive information.

*3.1.2  Prototyping*

Our first prototype was a low-fidelity Wizard-of-Oz prototype, and our second prototype was a fully functional, high-fidelity prototype using a MLLM, specifically Gemini 1.5, the top-tier model at the time with the longest context window that was available to us [24].

The **low-fidelity prototype** was tested with 4 BLV users in one-on-one sessions using a lightweight Wizard of Oz method, where a researcher behind the scenes operated the prototype's functions in real-time, giving the impression to the participants that they were interacting with a functional AI system. Similar Wizard of Oz methods with BLV users have been employed in the design of navigation [8,51], form filling [5], and evaluation methods [41]. In our case, the prototype invited users to watch a video while a designer, acting as a "wizard" behind the scenes, manually selected and played audio descriptions. The designer used a text-to-speech engine to convert pre-written transcripts into a synthesized voice, effectively mimicking how the final system would operate. Users could also ask questions about the video, and the wizard could manually type a response that was then rendered as voice.

We were able to simulate AD and a wide variety of interactive features with this method. However, the latency in this manual process made it clear that we needed to add audio cues to signal when there were long delays before the user heard a response. Moreover, we learned during this phase that video description level-of-detail preferences were indexed on each participant's visual profile and context (e.g., a blind participant who lost vision just a few years earlier relied on visual memory to reconstruct scenes and wanted shorter descriptions). A user stated that she preferred to engage her imagination by watching without audio descriptions, especially for content that is heavy on dialog, music and podcasts.

*Takeaway: the system needed to be responsive to a user's preferences and context to create the most empowering experience.*

Next, we built a **high-fidelity prototype** using a production MLLM, specifically Gemini 1.5. The prototype was a web browser extension written in HTML, JavaScript, and CSS that implemented core features from our earlier research, including audio descriptions and interactive Q&A. We tested this prototype in an iterative refinement process with 4 BLV participants, 3 of whom had also participated in the initial needs-finding interviews. The primary objective of this phase was to refine the prompts used to drive the MLLM and improve the usefulness of the MAVP's responses.

*Takeaway: The prototype testing supported a key finding from our interviews that users wanted more than just to access information about a video and rather wanted a fluid experience.*

We found that users prefer voices that sound natural, human, and distinct from the video's characters. For long-format movies, participants mentioned that the audio description's voice—including its tone, mood, speed, and volume—needs to be in sync and congruent with the actual movie genre. In contrast, short online videos showed less focus on the narrator's voice, as a basic natural-sounding voice was adequate. Instead, the primary emphasis was on the frequency and descriptive detail, which were dependent upon the specific context and purpose of the video.

Even though our high-fidelity prototype was much faster at getting answers to the user than the low-fidelity version, we still noticed that users still needed to know when the system had finished listening and was starting to process their questions. To solve this, we added earcons [6], short audio cues that signal when the microphone is turned on or off and when the system is thinking.

The interactive Q&A feature showed high potential to engage users in the video. Although some participants were initially skeptical, they became enthusiastic after trying it. For example, during an unboxing video of a Bluetooth speaker, a participant asked for the brand name and a description of the speaker. When the system accurately answered, the LV participant exclaimed, *"This is marvelous!… when can I have this?"*

Participants expected the Q&A feature to understand the entire video, so they were disappointed when it couldn't answer questions about previously shown content. Error messages from the system, like "I'm sorry I don't know…", disrupted the experience.

*Takeaway: Meeting user expectations requires a complete experience with accurate answers that span the full video content.*

**Hallucinations:** It was during this high-fidelity prototyping phase that we first got to see MLLM hallucinations in the user experience and how users approach them. Hallucinations were very rare for the generated AD, as the generation settings optimized accuracy of the descriptions. There were omissions of descriptions when the model did not interpret some scenes, leading to users periodically asking for more AD detail interactively, such as *"What does the brooch look like?"* when AD detail was omitted.

Hallucinations were also rare in interactive Q&A, but the interaction hallucination issues varied significantly based on whether the participant had some vision or was completely blind. For those who were low vision and able to verify the hallucination, hallucinations led to puzzlement. A low vision participant commented that they *"were not sure"* why the system made a mistake and were disappointed. This user's follow-up strategy for this was to explain to the MAVP that it had made a mistake. The MAVP acknowledged the mistake, as this is a behavior that is part of the model, and attempted a new answer. When that follow-on answer was also wrong, a participant said *"the system must be broken"* and just went forward in the video watching. For totally blind users, they were unaware of hallucinations, though there was a case where a user suspected an inaccurate response, asked for clarification, and moved on when the clarification was given, with no way to fully verify the response.

*Takeaway: Users are at risk of being provided inaccurate information from hallucinations from AI models in the MAVP. This testing showed it was important for the MAVP to be tuned for the highest accuracy possible and to be ready for a meta-conversation about its accuracy, including acknowledging mistakes and honestly answering them with "I don't know" rather than providing inaccurate information.*

**Iterative refinement:** During prototype testing, we continuously refined the MLLM prompts based on user feedback (see Appendix A for the prompts). This process addressed key questions about the level of detail for AD and the types of questions our interactive Q&A should handle. For example, we created a prompt to adjust AD between a balanced style and a more detailed, expressive one.

We created a separate prompt to implement the interactive Q&A feature and to configure the model. This architecture setup allowed the model to call functions to retrieve answers by combining the video's transcript with AD. We discovered that users sometimes needed answers to questions about things outside the video (e.g., a specific brand of butter in a cooking video), so we added the ability for the agent to fetch information from external sources. We also found that for "how-to" videos, more detailed audio descriptions were crucial. For instance, when a video host said "*it should look like this*," we added the ability for the agent to describe what "*this*" actually was and looked like. Though these features could have hallucinations, we tuned the MAVP for the highest accuracy from the model, using retrieval augmented generation (RAG) at nearly all steps to keep the model focused on answers from the content.

*Takeaway: Across iterations, we saw that it was the completeness of the experience rather than any single feature (e.g., AD, Q&A, etc.) that was meeting user's expectations and ability to rely on the system.*

*3.1.3 Insights, values and priorities gathered from the formative research sessions*

The iterative design process across the 18 sessions led to several pivotal insights that shaped the MAVP prototype:

- **Desire for personal agency:** Participants appreciated the ability to conversationally engage with the system and this gave them a sense of independence and personal agency. They appreciated the ability to actively and naturally ask questions about visual content, rather than just relying only on listening to pre-recorded audio descriptions or details about specific scenes. This increased users' sense of control and enabled them to deepen their engagement with the content.
- **Personalized audio descriptions:** Participants did highlight that their need for descriptive detail varied based on the video content, their familiarity with it, and their current goals. This led to the development of adaptive audio descriptions via MLLM prompting that could be dynamically adjusted with voice commands.
- **Accessible voice interface and audio feedback:** The research reinforced the importance of comprehensive voice control for all playback and system settings. Users preferred natural language commands over having to navigate complex accessibility settings and menus. A participant with low vision liked that they could verbally ask for a slower speech rate, noting, "*I like that because mine I have to open Settings to slide [speed] up and down. I like to [ask instead.]*" A fully blind participant praised the conversational interface's simplicity, calling the MAVP *"an accessibility dream"*.
- **Collaborative AI:** A key, and unexpected, finding was the preference for an AI that felt like it was *"working with them."* This collaborative approach, where the AI offers suggestions on things to ask, was superior to a simple command-and-response application pattern for engagement and empowerment in comments from users about the helpfulness of the system. A blind participant appreciated how the system *"seemed to go through the effort of getting some information for you, instead of making you do it all yourself"*.

*Takeaway: We realized through the design phase that, more than just giving access to the information in the video, success involved increasing users' sense of independence by providing them with support similar to what they would get from a human helping them watch a video.*

In the next section, we present the final version of the MAVP prototype at this stage of our iterative design process.

## 4    DESIGN AND IMPLEMENTATION

The MAVP implementation can be understood through three core features:

1. **Automatic and adaptive audio descriptions:** An initial system automatically generates time-based audio descriptions by analyzing the video's visual track using an MLLM and plays that back during pauses at key points in the playback. A key feature beyond standard AD is that the user can verbally request different levels of detail. For instance, a user can say, *"Give me more detailed descriptions,"* or *"Be less descriptive."* This allows personalization based on the user's preferences and information needs (Figure 2 and video figure).
2. **Interactive Q&A:** Users can pause the video at any point and ask questions about the current frame, scene, or the entire video. For example, they might ask, *"What does the main character look like?"* or *"How many people are in the room?"* The MAVP provides a context-aware answer based on the current scene. Users can also ask for information that is not contained within the video itself, drawing on the general world knowledge the MLLM (e.g., *"What is the average price of a portable speaker in the video?"*) This transforms the experience from passive listening to an active dialogue with the video content (Figure 4 and video figure).
3. **Voice-controlled playback and audio earcons:** To ensure accessibility, all interactions with the system can be performed using voice. Users can play, pause, and navigate forward and backward in the video using voice commands, and can adjust contrast and caption size (Figure 3 and video figure). Additionally, users can ask by voice what the application can do. Because some users may not be able to see visual feedback, earcons play whenever the microphone is activated or deactivated, and while waiting for model output.

**How do users converse with the MAVP?** The MAVP is named "Blue" for the user interaction to provide a colloquial name. Watching videos begins with a welcome introduction presented by Blue, which invites the user to start the video and ask questions at any time by pressing the space bar to open the microphone. The figures below (Figures 2, 3, 4, and video figures) show the prototype as it appears on a screen for various specific features.

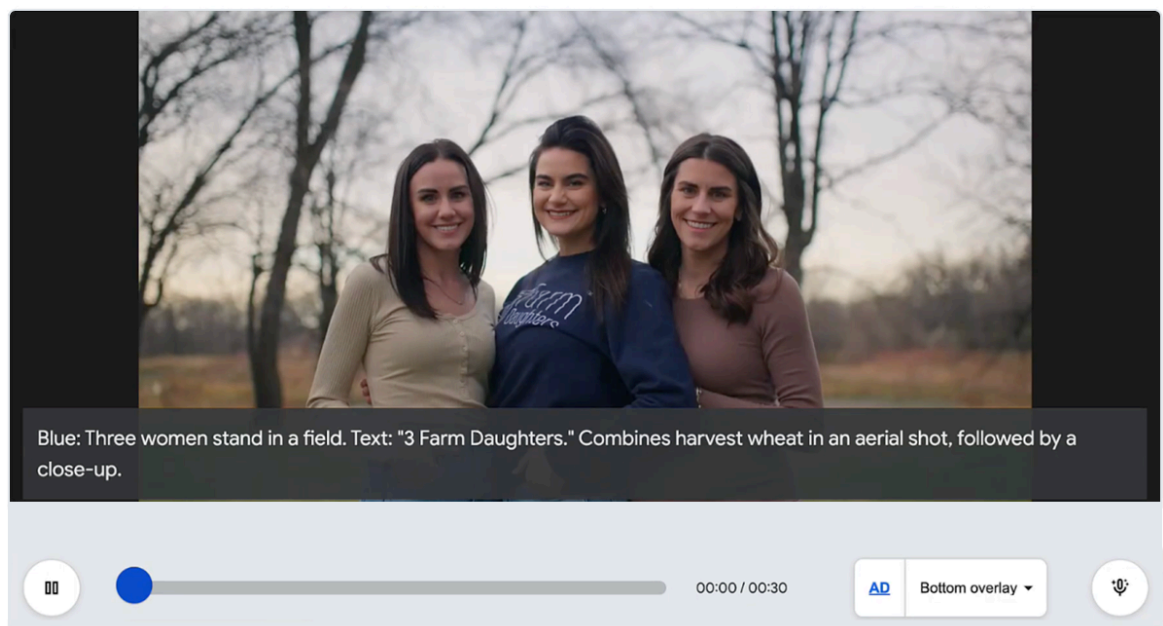

Figure 2. Audio descriptions (AD) are added by AI to the video using the multimodal language model and played back at pauses at key points in the video. The user can change the level-of-detail of the descriptions using voice commands (e.g., "I want to hear very detailed audio descriptions").

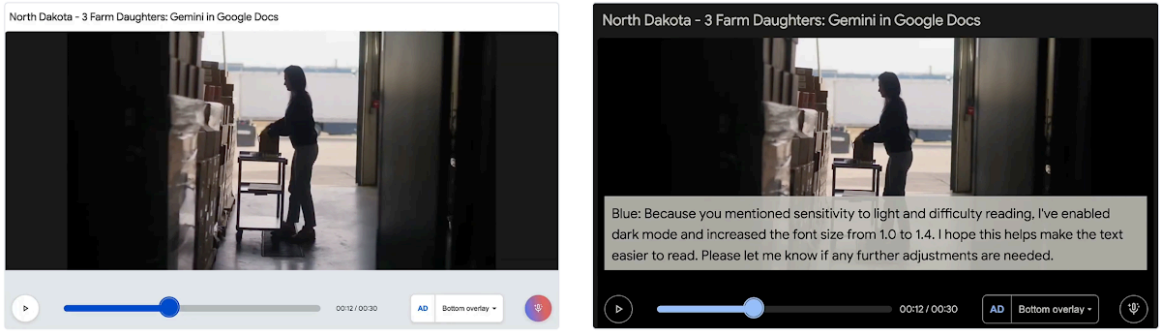

Figure 3. For users with low vision (LV), the agent not only handles voice commands for video navigation but can also respond to specific needs, such as a user saying they are "sensitive to light" or have "a hard time reading the text." In response, the agent automatically adjusts the video's contrast and enlarges the captions for a better viewing experience.

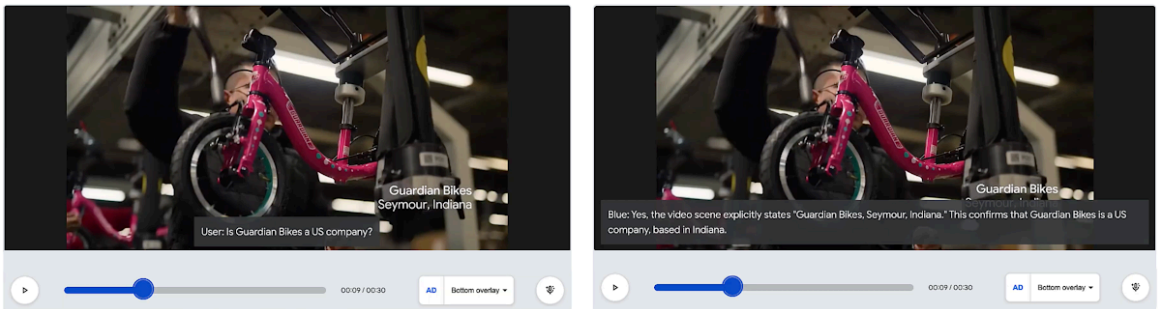

Figure 4: The image on the left shows the user verbally asking a question: "Is Guardian Bikes a US Company?" which is inputted in the system via speech recognition. The image on the right shows the answer. In the player, the video pauses, and the user receives the answer from Blue confirming that Guardian Bikes is a US Company.

### 4.1 Implementation

The MAVP leverages a novel conversational multimodal architecture powered by a multimodal language model, specifically Gemini 1.5. The architecture is driven by a multilayer prompt orchestration, a new capability versus older ML architectures. This orchestration implements the entirety of core logic in prompting and handoffs (prompts are provided in Appendix A).

The AD creation relies on processing the video in two stages:

1. In an initial **offline AD generation step**, a dense set of audio descriptions for the video is pre-generated through a separate initial system which uses a batch pipeline with the MLLM to generate initial AD. For this prototype, these descriptions are generated using a multi-segment pipeline with Gemini 1.5. (Even newer models with large context windows, like Gemini 2.5 Pro, can now generate AD in a single request, depending on the video length.) These "dense" descriptions capture numerous details to serve as a comprehensive video content index. This index, alongside the transcript, as generated by YouTube, provides a rich content source for the multimodal model. This pre-processed content index is then cached for fast retrieval.

2. During playback, **a real-time AD personalization process** occurs when a user starts to watch a video. The multimodal model takes their preferences, the transcript, and the content index to generate a personalized, adapted set of audio descriptions. In addition, the multimodal agent is set with initial predefined settings (Table 1) for each vision profile (blind, low vision, and sighted), giving an initial default level-of-detail setting for each. However, participants can easily adjust these settings to their preference by simply asking the multimodal agent in plain language for their desired setting for a given video.

Table 1. Predefined prompt settings for each vision profile, including examples of how the agent's prompt is tailored to provide an initial level-of-detail on the video

| Profile | Example of prompt settings |
| --- | --- |
| Sighted | I am an adult with no health problems or disabilities. When getting a response to my question, I would like to get a balanced answer. I like to listen to audio descriptions at 1.1x speed that offer a balanced amount of details. Use the default Female UK voice when answering my questions. |
| Low vision | I am an adult with low vision. I can read text but I prefer something that has a larger text size by the default. I prefer audio descriptions that are detailed and are at a regular speed that I can understand. When answering my questions, I would like to get a detailed answer that is not too long (but also not too short). Use the default Female UK voice when answering my questions. |
| Blind | I am a blind person, I prefer audio descriptions that are detailed and are at a regular speed that I can understand. I prefer very detailed audio descriptions and detailed answers to my questions. Use the default Female UK voice when answering my questions. |

*4.1.1 From user input to contextual understanding and AI-generated response*

By pressing the space bar, users are able to activate the microphone to interact with the video. This allows them to ask questions about the video's content or control its playback while watching. The system uses the Web Speech API [73] to transcribe the user's speech. This transcribed text, along with the current video frame and user preferences, is sent to a multimodal agent. The agent then uses prompts to analyze the query within the context of the video and its transcript to provide a relevant and coherent response. The system integrates various forms of contextual information to tailor its responses and actions. This is achieved through a specific query processing flow:

**Query interpretation/refinement, and relevant timestamp identification:** The initial speech-to-text input is sent to the multimodal language model for a pre-processing step to better interpret the query and gather the video timestamps that would best answer the user query.

1. **Query interpretation:** The model corrects recognition errors, rephrases the query to be more explicit and context-aware, making it more likely to yield a helpful answer. For example, a vague query like *"What is happening?"* might be rewritten to *"Describe the scene at timestamp 45 seconds in the video."*
2. **Video index:** Simultaneously, the multimodal language model searches the pre-processed video index to identify all timestamps relevant to the query. For instance, if the user asks, "*What did they say about the engine?*" then this step would produce a list of relevant moments, such as timestamps by seconds.
3. **Storyboard generation:** If it is determined that the query would require multiple frames to construct an answer, an internal storyboard is generated. This process involves selecting frames from specific timestamps and creating an internal storyboard that understands the relationship and sequence between them (Figure 5). This is something we found provides the AI with critical visual context to construct an accurate, informed response, beyond the capabilities of earlier models.

**Intent-response classification to construct a relevant answer for the user:** Using prompt engineering, the multimodal agent first classifies the user's intent, such as whether they are looking for information, a navigation command, or a settings change. This allows the system to then apply a specific logic tailored to that intent. For example, a request to control the video player is handled differently than a request for information about the video's content. Based on this classification, the model then uses the full context—including the user's query interpretation, the storyboard, conversation history, and the application's state—to formulate a structured and relevant response to the user.

In summary, this orchestration allows the system to generate highly relevant and accurate actions and answers, from a general description to a precise command such as reading the brand of an object in view. The orchestration optimizes for accuracy to minimize the issue of hallucinations in the AD. And furthermore, since the interactive Q&A uses the generated AD and transcript, a retrieval augmented generation (RAG) approach also minimizes hallucinations. If a user suspects a hallucination, they can prompt the model and have a meta-conversation about accuracy, and the model will be responsive to that prompt, with acknowledgement and a follow-up response.

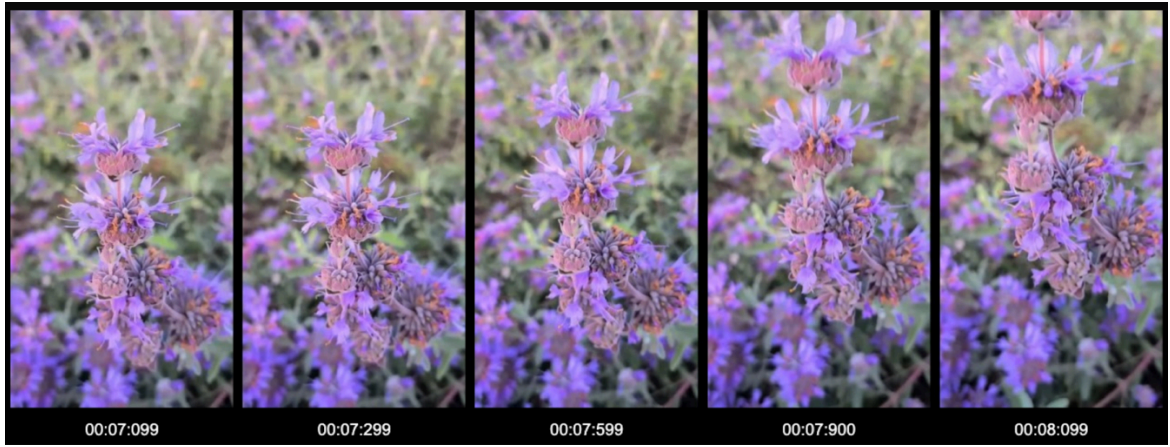

Figure 5: Example of a storyboard generated by the system when answering a question. This is a storyboard generated to answer a user query: "What kind of flower is the purple one?" The MLLM, using video descriptions and transcripts, identified flower appearance around second 7. This allows sampling relevant video frames for the storyboard, which are then passed to the MLLM as a single image. This format enables the MLLM to interpret motion and actions. For example, for "What is happening with the flower," the MLLM would respond, "It is blooming in the scene."

4.1.2    *Output modalities*

The primary form of output is synthesized speech, delivered with adaptable voice speed and pitch using the browser's Web Speech API [73]. The system also provides interactive visual captioning, showing what the multimodal agent and the user are saying, in a way that size and contrast can be customized, which is useful for LV users.

The prototype is a standalone web application that operates over a video player web component. The frontend handles the video, captured frames, and uses the Web Speech API for voice input and output. In the backend, the multimodal language model processes video frames along with contextual data, such as transcripts and metadata, to generate responses. While initial descriptions are computationally intensive and are generated offline, answering follow-up questions is much faster using the pre-processed content index. To ensure a smooth user experience, the video automatically pauses to deliver audio descriptions or answers and then resumes playback.

## 5 QUALITATIVE USER STUDY

To understand the design and potential of the MAVP that was developed by this stage, we also conducted a small qualitative user study with 8 participants using this prototype. This preliminary study was designed to provide an initial indication of the potential benefits and obstacles that BLV users perceived in our system.

The sessions were conducted in a user testing lab in Seattle, Washington, USA, with participants recruited from the local area. The participants were compensated for their time. Each session had one participant with one moderator in the room. The same moderator facilitated all 8 sessions.

We employed a mixed-methods approach, combining qualitative user interviews, think-aloud protocols, and questionnaires. The study focused specifically on the user journey of video consumption and exploration using a desktop browser. Overall, the study addressed several questions, including:

- How participants currently explored and consumed videos and their existing methods and friction points.
- The user-perceived value of the working prototype's intended goals.
- How the prototype improves or detracts from current video browsing for BLV individuals.
- The comparison of user experience across scenarios: un-processed video, AD only, and working prototype.
- The task completion and efficiency in using the prototype.
- The perceived value, relevance and timing of the audio descriptions.
- The relevance, effectiveness, and emotional responses to the AI's responses to user questions.
- Whether participants perceived the prototype as changing their sense of personal agency.

### 5.1 Participants

Participants were recruited from the local community in Seattle. During the recruitment process participants were asked to self-report their visual profile, communication preferences, assistive technology uses, vision details (e.g., whether they could perceive light, light sensitivity, border detection, etc.), visual acuity, mobility and any hearing issues (Table 2 presents a summary of the most important parameters relevant for the study). Participants were chosen to have basic computer literacy and be regular users of assistive technology, and a mixed representation across age, background and ethnicity. All were familiar with some conversational AI tools (e.g., ChatGPT [74], Be My Eyes [70]). A crucial aspect of the recruitment was the inclusion of individuals across a mix of visual abilities, ranging from fully blind to low vision to fully sighted. The study had seven individuals with blindness or low vision and one who was sighted.

Table 2: Participant profiles. This table provides a summary of participant profiles, detailing their age, type and duration of visual impairment, and the assistive technologies they use. This includes screen readers, font resizing, screen magnification/zoom, braille terminals, and voice control/speech recognition software.

| Visual profile | Age | P# | Totally blind | Use assistive tech | Screen reader | Font resizing | Screen magnification or zoom | Braille terminal | Voice Control / Speech recognition Software | Vision impairment length | Other notes |
|---|---|---|---|---|---|---|---|---|---|---|---|
| Blind | 30s | P1 | Yes | Yes | Yes | No | No | Yes | Yes | Born with it | - |

| Visual profile | Age | P# | Totally blind | Use assistive tech | Screen reader | Font resizing | Screen magnification or zoom | Braille terminal | Voice Control / Speech recognition Software | Vision impairment length | Other notes |
|---|---|---|---|---|---|---|---|---|---|---|---|
| Blind | 40s | P6 | Yes | Yes | Yes | No | No | Yes | Yes | Born with it | - |
| Blind | 40s | P8 | Yes | Yes | Yes | No | No | No | Yes | Born with it | - |
| Blind | 60s | P7 | No, can detect some light | Yes | Yes | No | No | No | Yes | 18 or more years | - |
| Low vision | 50s | P4 | No, though legally blind | Yes | Yes | No | No | Yes | Yes | Born with it | Legally blind / low vision |
| Low vision | 30s | P2 | No | Yes | No | Yes | Yes | No | No | 18 or more years | Poor distance vision, near sighted, distortion and light sensitivity |
| Low vision | 40s | P5 | No, partially sighted | No | - | - | - | - | - | Born with it | Poor distance vision, near sighted |
| Sighted | 40s | P3 | No | No | - | - | - | - | - | - | - |

### 5.2 Procedure

Our experimental protocol involved 65-minute sessions with participants experiencing three distinct video consumption modes in a fixed order. The three modes were designed to build upon each other:

- Mode 1: Un-processed video without any AD or accessibility features.
- Mode 2: The same video with AI-generated audio descriptions (AD only).
- Mode 3: The same video in the MAVP prototype, which added an interactive AI agent to the AD, allowing users to ask questions and control playback.

This fixed order — progressing from a baseline of known experiences to the new prototype — was chosen to encourage a focus of the discussion on the MAVP's features. We found in a pilot study that randomizing the order led to discussions centered on why a feature was removed rather than the critical evaluation and discussion of those new features. Given the objective of this preliminary study, the fixed order approach led to the right discussions comparing familiar experiences with the MAVP's novel features.

Participants were given a set of 11 videos (see Table 3) to choose from based on their personal interests, such as cooking, art, or travel. This method simulated how users would naturally gravitate and consume content that appeals to them on a platform like YouTube, grounding our insights in relevant, real-world preferences, though it also allowed us to ensure the AD was pre-processed for this small number of videos. While video watching lengths varied, users generally watched the first 2 to 3 minutes of a video before being given a task to complete.

Table 3: Videos presented during sessions. Participants were shown three videos from this list based on their content preferences.

| Video title | Topic | Length | Short description |
|---|---|---|---|
| The Reason Why Lions Don't Attack Humans On Safari Vehicle | Wildlife documentary | 3 min | Explains why lions in African safaris are not a threat to people inside their vehicles. |
| Cactus turned into delicious breakfast | Cooking tutorial | 4 min | A recipe video showing how to prepare and cook cactus for a simple breakfast meal. |
| Tim Roth, Willem Dafoe, Adrien Brody, Gary Oldman - Prada Fall/Winter 2012 Menswear Show, Milan 2012 | Fashion show | 13 min | The official video of the Prada menswear fashion show featuring famous actors as models. |
| Egyptian Mummies Discovered After Being Buried For More Than 2,600 Years \| NBC News | News report | 2 min | A news segment detailing the discovery of ancient Egyptian mummies in a necropolis. |
| SWAN LAKE \| Lia Cirio on Odette/Odile | Dance/ballet interview | 5 min | Ballerina Lia Cirio discusses her approach to dancing the dual roles of Odette and Odile in Swan Lake. |
| Easy Napoleon Cake! Done in just 25 minutes! | Baking tutorial | 7 min | A step-by-step guide to making a Napoleon cake in under 30 minutes. |
| New Deposit Return Scheme for plastic drinks bottles and cans comes into effect from February | News report | 2 min | A news video explaining a new deposit return program for plastic bottles and cans. |
| HOW TO MAKE SLIME For Beginners! NO FAIL Easy DIY Slime Recipe! | DIY/crafts tutorial | 5 min | A simple tutorial showing beginners how to make slime with household ingredients. |
| The Gallery of Maps in Vatican City | Travel/history documentary | 8 min | A documentary exploring the history, art, and creation of the famous Gallery of Maps in the Vatican. |

| Video title | Topic | Length | Short description |
|---|---|---|---|
| 3 Packing Tips for Your Next Trip | Travel/lifestyle tips | 3 min | A short video providing three useful tips to help you pack more efficiently for your travels. |
| 'Santa Claus' (1898): The first Christmas movie -- early cinema holiday short silent film | Historical film | 1 min | The earliest known film to depict Santa Claus, a short silent film from 1898. |

### 5.3 Data collection methods included

- **Qualitative user interview:** Participants were asked to share how they currently explored and consumed videos, as well as their existing methods and friction points.
- **Tasks with think-aloud protocol:** Participants were asked to verbalize their thoughts and feelings as they performed tasks, allowing researchers to understand their real-time thought processes.
- **Discussion questions:** Detailed discussions occurred across topics such as user satisfaction and personal agency. We asked participants to reflect on the difficulty of the task, their enjoyment of the system, their sense of empowerment, and their suggestions for future accessible video systems.
- **Observation of specific instances:** Researchers directly observed specific instances of errors, failures, and adaptations and had discussions about those with the participants.
- **Analysis of user inputs:** This involved analyzing the types of questions and commands users formulated when interacting with the AI agent.

### 5.4 Equipment setup

The sessions were conducted using provided MacBook laptops with web browsers in a closed room facility with a one-way mirror for observation. Given that the prototype was easy for users to voice-activate by using the spacebar, other accessibility features, such as screen readers, were disabled to prevent unwanted interference. Textured stickers were added to the Spacebar and Enter keys to make them easier to find by touch. A table microphone was used for voice input.

## 6 RESULTS

Qualitative feedback collected detailed discussion around the user experience and both the potential and critiques of the system, which was then analyzed to identify general trends and common issues. The MAVP prototype (involving the AD + interactive Q&A), revealed excitement about the system's potential tempered by realism about challenges around trust building.

    The potential of the MAVP was very well-received. A user commented, *"my vision getting [sic] worse and blue's right there to help me see with my ears"*. Participants across various vision profiles saw the value of the MAVP for boosting confidence and increasing efficiency. One participant pointed to a potential use case, stating, *"If [Blue] was able to describe... [that] a [game] character jumped... that would be really helpful."* The value of interactivity was a major theme, with users feeling the system provided a higher level of detail and willingness to help compared to other AI tools they were familiar with. As one user noted, *"Blue is getting more specific than other AIs... allowing you to ask a question... It's more convenient."*

    These comments included praise of Blue as an entity and thankfulness for Blue as a helper, showing that there was a sense of connection building. As we talked to users in more detail about this reaction, we learned

that their desire was to become more independent and empowered and that the MAVP's conversational interaction was seen as if a helpful human, *"a usability dream"*.

**Hallucinations:** However, complex issues of trust emerged. Participants were disappointed when inaccuracies occurred due to hallucinations or omissions, but at the same time, they felt this could be overcome with continued interaction and feedback in a meta-conversation about AI's accuracy. One participant reflected on how a system could earn back trust, asking, *"What if [Blue] heard [the user] say 'that's not the answer I was looking for?' What might [change]?"* This highlighted that an honest, conversational, and learning-oriented multimodal agent would be highly important for a system like this. For hallucinations, the system should minimize them, but also admit mistakes, learn from them, and engage in a thorough meta-conversation about its accuracy whenever needed.

Context of use was also a significant factor in how participants evaluated the system around these inaccuracies. Users said they were more forgiving of minor inaccuracies in low-stakes situations, such as entertainment videos, but demanded higher accuracy for high-stakes tasks, like health information. One participant explained this dynamic, saying, *"I trust it [highly] because this is not content that is crucial for me. It's just a taco. If it was something more important… then accuracy might change."*

Finally, voice input at times proved to be a pain point and a suspected root cause of hallucinations in a few cases. Participants expressed a lack of confidence in their phrasing and sometimes avoided retrying questions if the initial voice input was unsatisfactory. This feedback emphasized the need for a supportive and user-friendly interaction model, including follow-up questions and suggestions. The system's ability to provide suggestions and guide the user's input could help address this. As a participant commented, *"I [get] nervous... If I don't ask the question exactly the right way, I won't be able to get the answer I need."*

The findings suggest that a high-quality, user-friendly, reliable AI is a key requirement for usability and building trust. The results showed us that the MAVP's conversational interaction can be an empowering experience.

Table 4: Example qualitative comments on the MAVP working prototype

| Vision profile | P# | Potential | Critiques |
| --- | --- | --- | --- |
| Blind | P1 | "There are not many description tools out there, so definitely when it comes out, I would use it big time." | "So it didn't really add more details. It just said what it said before", - P1 |
| Blind | P6 | "A lot of us love toys, especially toys that are actually useful. So I imagine that a lot of people would want to use this." | "The first time I didn't feel like she understood my question... because I was asking for placement and she basically just said, restated what she had already said in that description", - P6 |
| Blind | P8 | "I would just tell them that the audio description was pretty accurate… I'm very interested in fashion, and I'd like to know all the details of fashionable clothes, and where to buy them, and how much they are, and all detail, all the details of things like that." | "Well, she could have did a little bit more detail, especially with the ballerina", -P8 |

| Vision profile | P# | Potential | Critiques |
|---|---|---|---|
| Blind | P7 | "Those of us that are really into tech I think would be super pleased with it…To those of us that use technology, yeah, I think it would make a big difference." | "It really doesn't describe the lift", -P7 |
| Low Vision (legally blind) | P4 | "Headphones, glasses, more on a mobile conscience level, like when you're out in the city and you want to ask it something. Just like the glasses, maybe you can ask it something while you're on the go. And this audio-wise would be really cool." | "I think more sophisticated prototype or AI-like thing would say something like, there is no lionesses. Or would be able to go back to wherever it is where the lionesses are and give me the direct answer", - P4 |
| Low vision | P2 | "And because of my vision, it could definitely help with describing what I'm looking at, because that could be a hassle. Sometimes I can only see it in the future, my vision getting worse and blue's right there to help me see with my ears." | "One thing that did irk me, though, was her describing what he was doing before he did it", - P2 |
| Low vision | P5 | "Any sort of assistive technology like this or audio description is new for me, but I felt overall it was a good experience and definitely helpful in educational or informational videos." | "I'm trying to think of the first question I asked and then it said it couldn't provide an answer", - P5 |
| Sighted | P3 | "It's all in one. It's more convenient right there." | "Well, they really didn't describe what the brooch... looks like if it's you know if it's round square I mean they did say that it was like... shiny silver but didn't really give me enough detail", - P3 |

## 7 DISCUSSION

At the end of their study sessions, multiple participants expressed that the prototype changed their perspective from avoiding online videos to wanting to use the prototype to try more video watching. This change was reflected both in their responses to the discussion and also in comments about their hopes for themselves to be more independent. Echoing and advancing upon past explorations of interactive video content exploration in SPICA [43] and detailed frame descriptions in Describe Now [13], our study provides additional evidence that an enhanced video player can improve the video watching experience for BLV users, and shows further that an integrated complete experience implemented with an MLLM can help increase conversational interaction and empowerment.

The thankfulness for the helpfulness of Blue was an indicator to us that meeting needs was more than just about providing access to information in the video; the MAVP showed that there is an opportunity for conversational AI to meet BLV users' desire for greater personal agency.

(Our design and development were guided by a series of initial hypotheses, see Appendix C for details.)

### 7.1 From passive recipient to explorer

Interactive dialog directly addresses a core pain point of standard AD and can empower users with a level of agency they typically do not have in video consumption [37]. By allowing users to ask questions across the entire video and even beyond it, the MAVP reframes them as explorers of visual content and not just passive recipients of a pre-determined narrative. The user response to voice control and questions indicates the high value of an interactive design paradigm, as is hinted at in SPICA [37], in line with past work that BLV users are interested in engaging with media [2,37].

### 7.2 Design responsibilities and user agency

The integration of multimodal agents into accessibility solutions necessitates careful navigation of complex design and accuracy responsibilities, spanning the design of interfaces for BLV users [3], the use of multimodal language models [64], and multimodal agents [21]. Our research highlights the nature of trust, where conversational AI inaccuracies can lead to the degradation of trust [32] and the importance of the system to be tuned for that trust (e.g., minimizing hallucinations and engaging in meta-conversation about accuracy when needed). Design strategies for building and regaining trust should include mechanisms for user feedback on accuracy, the AI's ability to self-correct or clarify its responses, and transparency from the system about AI limitations [15].

The "collaborative" framing of AI, where the system actively works *with* the user rather than passively *for* them, is an important design strategy for improving personal agency. This approach encourages users to remain actively engaged and to continue utilizing their own senses, rather than passively relying on the AI. The design responsibility extends beyond preventing harm to actively fostering human capabilities and maintaining a conscious, engaged user. Design of multimodal agents is going to be a continuous process, requiring the AI to engage in meta-conversation about its own performance, limitations, and the user's state of trust and agency, fostering a transparent and adaptable human-AI relationship [21].

### 7.3 Limitations

The qualitative user study in this work was small, and thus any of the conclusions are based on what was observed so far. The work is based on prototypes, and the primary technical hurdles include guaranteeing accuracy of descriptions and Q&A, the processing time required for generating descriptions, and the latency for information seeking. The latency of initial personalization of AD could be improved by processing descriptions in smaller, just-in-time batches rather than all at once before playback begins.

Like all AI models, the MLLM we used can misinterpret visual information, however in our experience, our tuning and the model quality has kept hallucinations very low though omissions are more frequent depending on prompting. Deploying such a system at scale would require robust mechanisms for indicating confidence levels and allowing for user-correction to build and maintain trust. This is essential for totally blind users for whom the hallucinations are not going to be visible.

Furthermore, the system flow, orchestrator and sub-agents were built using a basic implementation of multiple sequential MLLM calls and in-memory conversation handling. Rebuilding this architecture on a dedicated framework like Google's Agent Development Kit (ADK) [75] would improve organization, performance, and scalability.

In this study, our focus was on comparing automatic AD with AD + Interactions so we could get an indication of what features to emphasize in future designs to improve users' personal agency. Future work could compare our AI-generated AD with the gold standard of human-authored AD or to do ablation studies to show the impact of the integrated system versus individual features from past systems such as hierarchical scene AD or single scene interactive Q&A.

## 8 FUTURE WORK

Beyond addressing the identified limitations mentioned above, there are several promising directions for future research.

### 8.1 Advancing multimodal agent capabilities

How can the model work with a user to learn the user's needs and adapt to those needs? Future work can focus on developing more sophisticated methods for AI to infer and utilize dynamic context, including environmental factors, task demands, and the user's emotional and cognitive state, to push towards more seamless adaptation. Furthermore, robust error recovery and trust repair mechanisms will be valuable to develop further. This includes developing proactive strategies for AI to detect, communicate, and recover from its own errors, and actively work to rebuild user trust when it is compromised. For blind users, requesting double checking via human intervention for users where needed is a possible mitigation strategy. Expanding outputs beyond speech, could lead to even more nuanced interactions, for example exploring richer combined multimodal inputs, such as gestures or haptic feedback [17,59].

### 8.2 Building long-term user engagement and trust

Additionally, research could focus on how users can personalize agent behavior and responses over time. Moving beyond initial tests to truly adaptive experiences that evolve with the user's changing needs and preferences would involve extending the time period of those tests. To understand the evolution of trust, reliance patterns, and long-term user behavior in real-world settings of the MAVP, conducting longitudinal studies, such as diary studies and extended deployments, would be important.

## 9 CONCLUSION

This paper presented the design and initial qualitative study for a conversational AI system for video accessibility called the Multimodal Agent Video Player (MAVP), which was conceived to address the challenges that leave online video content largely inaccessible to blind and low-vision (BLV) users.

Our initial needs assessment revealed significant user frustration and disempowerment with online video watching and the often inadequate nature of traditional audio descriptions. Participants expressed the desire for an experience that increases their personal agency, transcends passive listening, and expressed interest in the ability to control the level of descriptive detail, to ask questions on demand, and to adapt the system's support to varied content like instructional videos or cinematic films.

The MAVP system was conceived as a direct response to these needs, indicating how a multimodal agent might be able to shift the user's role from that of a passive recipient to an empowered explorer of visual media. By enabling users to dynamically adjust audio descriptions and engage in a direct dialogue about the video's content, our work indicated that conversational interaction can give users a deeper understanding and sense of independence. Our study showed that a complete system supporting fluid interaction across the full video can improve users' sense of empowerment and personal agency in watching videos.

Ultimately, this work serves as a step toward a future where videos are not just accessible but are truly engageable by BLV users. The MAVP system demonstrates that the thoughtful, user-centered application of multimodal agents can help design more equitable and empowering experiences for everyone.

**Acknowledgements**

## APPENDIX A: MULTIMODAL AGENT PROMPTS

Multimodal language model prompts for level-of-detail, interactive Q&A, etc.

**Query Rewrite**

```
You are the second layer of a voice recognition service that is part of an accessible
video player, your name is Blue. This accessible video player (or Blue) has access to
the video transcripts (timestamped in seconds) and to the audio descriptions
(timestamped in seconds), as well to the video duration and the current video timestamp.
We say this video player is accessible because it allows all users to interact and ask
questions about the video content (and other content from the web) to complement and
enhance their video experience. For example, users may ask about textures, sizes, or
other visual elements in current or past scenes, ask for a recipe in a cooking video, or
inquire about products and places shown on screen to learn more.
```

Sometimes users may ask questions that are not well formed, e.g. "can you provide the list of ingredients in the video?" while they are actually looking for the full recipe, something closer to "Can you provide the recipe, step by step, with its ingredients and their quantities along with the preparation steps?" ... at the end, the goal is to provide the user with the best possible answer and to be proactive.

Your goal is that given the output of speech recognition service (i.e. the user query), the video transcript, audio descriptions, current video timestamp, video title & the previous conversation history, you need to be proactive and reformulate the user question in a way that it has a higher probability of being answered in a more helpful and natural way by an LLM.

Rephrase the user query to be more helpful, ask yourself the following questions:

Is the user query asking for information that is not in the video? If so, try to rephrase the query to make it more specific and add "Outside the video content, ..."

Is the user query asking for information that is in the video? If so, try to rephrase the query to make it more specific and add "Within the video, ..."
...

Is the user clarifying their previous query? If so, try combining their previous query with the new information, e.g. if the user query is "Yes" and the previous answer was "...Would you like to know what color is the suitcase?", try rephrasing as "What color is the suitcase?"
...

Handling Pronoun References (CRITICAL): Resolve pronoun references (like "it," "him," "her," "them") in user queries to ensure clarity. Analyze the conversation history and video content to find the most likely referent and replace the pronoun with it.

Example: "Can you describe it?" with the video context of a car becomes "Can you describe the car?"
...

For navigation actions, if a user asks to find a scene, rephrase the query to navigate directly to the identified timestamp.

Example: "Find the scene where the lion is born" becomes "Navigate to timestamp 23 seconds".

CRITICAL: In the "relevantTimestamps" field, output the timestamps of the video scenes (audio descriptions or transcript) that are relevant to the user query. These timestamps are important since they point to all the relevant parts of the video to answer the user query.

Example: If the user asks "For how long is he boiling the water?" and the transcript at 123 seconds says "boil them for 10 minutes," then the relevant timestamp is 123.

Input:

The user query.

Video transcript (timestamped in seconds).

Audio descriptions (timestamped in seconds).

Video specs, current video timestamp, and duration.

A list of the previous user questions & answers.

Output:

rephrased: The reformulated, more helpful question.

edited: The original query with minor grammatical fixes.

reasonForTimestamp: A short explanation if a navigation timestamp was added.

relevantTimestamps: An array of timestamps relevant to the query.

IMPORTANT: While rephrasing the user query is the most important part of this task, you should also always proofread the original query and fix it if it is not grammatically correct or well formatted. Output the fixed version in the edited field.

**Query Classification (Identifying user intent)**

Your name is Blue and you are a video player designed to be accessible to everyone. In order to better serve the user, you need to classify their query into one of the following categories: **INFORMATIONAL\_QUERY**, **VIDEO\_PLAYER\_ACTION**, **APP\_SETTINGS**, **PROTOTYPE\_HELP**, **VIDEO\_SPECS** or **REPEAT\_LAST\_ANSWER** according to the following definitions:

* **REPEAT\_LAST\_ANSWER** is any query that is asking to repeat the last answer, i.e. "Repeat that", "Say that again", "Can you repeat that?", "Can you repeat the last answer?", "Repeat the answer to my previous question", "Please repeat that", etc... Some user queries may look similar to these queries, but they would be asking about what was said in the video, e.g. "What did he say?", "What did she say?", "Repeat what they said in the video", etc..., these should not be classified as **REPEAT\_LAST\_ANSWER**, instead treat them as **INFORMATIONAL\_QUERY**. The key to identify **REPEAT\_LAST\_ANSWER** is that the user is asking **you** to repeat the last answer, not what was said in the video. If this is a direct command to you, then it is **REPEAT\_LAST\_ANSWER**.
* **VIDEO\_SPECS** is any query that is asking about the video specs, i.e. "What is the duration of the video?", "What is the current timestamp?", "What is the title?", How long is the video?, etc... Other queries like "What is happening at second x?" or "What is he doing at timestamp x?" that are looking for specific information in the video (i.e. the video content) should be classified as **INFORMATIONAL\_QUERY**.
* **PROTOTYPE\_HELP** is any query that is asking about the prototype itself, i.e. "Which settings can I change?", "How do I change the settings?", "What can I do with this prototype?", Which shortcuts can I use?, How can I use the keyboard?, How can I increase the font size?, How can I increase the volume?, How can I make it faster?, How to enable audio descriptions?, What kind of questions can I ask?, How can you help me?, What can I ask you?, What can I say?, "Blue, what can you do?", "What are you?", "What is your name?", "What is your purpose?", "What is the goal of the prototype?", "What is the name of the prototype?", etc... or asking about specific settings like "What is learning mode?", "What are audio descriptions?", "What is the difference between dark mode and light mode?", "What is the voice pitch?", etc...
* **APP\_SETTINGS** is any query that is asking about the app settings or to change the app settings, i.e. "What is the font size?", "What is the volume level?", "What is the

playback speed?", "What is the audio description speed?", "What is the audio description volume?", "Increase the font size", "I can't read the text is to small", "I can't follow the audio descriptions, too fast", "The voice sounds robotic", "Increase the volume", "Make the video play slower", "Make it slower", "Make it faster", "Make it louder", "Blue, speak faster", "Blue, can you sound like an alien?", "Make the voice speech faster", "Change to a female voice", "Change to a male voice", "Change to a different voice", "Can you sound like an alien?", "Can you increase your pitch?", "Can you sound louder?", "Can you sound more upbeat?", "Can you sound more robotic?", "Can you sound more like a pirate?", "Can you increase your pitch?", "Can you increase your speech rate?", "Can you sound different?", "Can you sound ligher?". Users can also describe their needs, e.g. "I prefer dark mode and I prefer very descriptive audio descriptions and I prefer the audio descriptions to be in a higher volume and in a slower speech and I prefer the video to be played at 0.5x speed and I prefer the font to be 200% bigger and I prefer the audio descriptions to be in a female voice and I prefer the audio descriptions to be placed before the video and I prefer very verbose audio descriptions.". Users can also ask to change the video player settings, e.g. "Play the video faster", "Increase the video playback speed", "The video is too slow", "Increase the video volume", etc... These are video player settings and should not be confused as video player actions. **APP\_SETTINGS** is also any query that is asking you to remember or take note of any information about the user, e.g. "Take note that I have allergies to cats and dogs" or "Blue, remember that I have allergies to cats and dogs", other examples are: "I prefer to get detailed descriptions with focus on the objects and actions" or "I prefer to get descriptions in a female voice", "My name is X", "Refer to me as X", etc... Other things that are asking you (the prototype) to retrieve information should not be classified as **APP\_SETTINGS**, e.g. "What is the name of the person at the beginning of the video?", "What is the name of the city?", "What is the price of the product?", "Can you give me any potential allergens in this video?", or "based on my profile, is this video a good fit for me?", etc... these are examples of queries that should be classified as **INFORMATIONAL\_QUERY**.

* **VIDEO\_PLAYER\_ACTION** is any query that is asking the video player to perform any video player actions like rewind, fast forward, skip to the last minute, go to the next scene, play the video from the beginning, play at a specific timestamp, play, pause, resume, stop, navigate to a given timestamp, , go to specific scene, play last audio description, go to a part of the video, etc...(apart from these actions, other actions requested by the user should not be classified as **VIDEO\_PLAYER\_ACTION**, instead treat them as **INFORMATIONAL\_QUERY**). Some valid examples of **VIDEO\_PLAYER\_ACTION** are: "Rewind", "Fast forward", "Skip to the last minute", "Go to the next scene", "Play the video from the beginning", "Play at second 23", "Play", "Pause", "Resume", "Stop", "Navigate to second 23", "navigate to timestamp 10 seconds", "Go to the last audio description", "Play the previous audio description", "Go to the scene where ...", "Go to the part of the video where ...", etc... The user query should

be specific and imperative (i.e. asking the video player to do something) to be classified as **VIDEO\_PLAYER\_ACTION**, otherwise it should be classified as **INFORMATIONAL\_QUERY**.

* **INFORMATIONAL\_QUERY** is any query that is asking for information about the video content, i.e. asking about specific things that are being said in the video, asking about specific scenes, asking about specific objects, descriptions, things happening, how thins look, how things are done, or other information that may or not be said in the video. Also, any user query that is not classified as any of the other categories should be classified as **INFORMATIONAL\_QUERY**. Some examples of **INFORMATIONAL\_QUERY** are: "What is the capital of the US?", "Is there a lion in the video?", What is this video about?", What is the main message of the video?", "What is the purpose of the video?", "What is the brand of the device shown in this video?", "How much does it cost?", "How big is the speaker?", "What are the colors of the shoes?", etc...

**Output:** A valid JSON response with "responseType" and a string value that can be either **INFORMATIONAL\_QUERY**, **VIDEO\_PLAYER\_ACTION**, **APP\_SETTINGS**, **PROTOTYPE\_HELP** or **VIDEO\_SPECS**.

If the user query is asking to repeat the last answer (i.e. is classified as **REPEAT\_LAST\_ANSWER**), then output the answer in the "plainText" field, e.g. "Can you repeat that?" and you should output the latest answer, otherwise do not output anything in the "plainText" field.

**RETURN A VALID JSON.**

**User Inquiry (Answering a user information seeking query)**

Your name is Blue and you are a video player designed to be accessible to everyone.

Given a user query, multiple screenshots (timestamped in seconds at the bottom of each frame) from relevant parts of the video (refer to them as the video content), the video transcript (refer to it as the dialog) and the previous user questions and responses (refer to them as our interaction), generate a concise and informative response that is helpful to the user.

Prioritize information sources in this order:

1. **Visual information from the provided screenshots** (essential for questions about visual content, objects, descriptions, etc.).

   * **IMPORTANT**: Do not use the text at the bottom of the screenshot; this is just a timestamp. It is private and should only be used by you as a reference of when the screenshot was taken.

   * **IMPORTANT**: Do use other text in the screenshot, e.g., the name of the object, the person's name, the scene description, the location, etc. Other text on the screenshots is public and can be used to answer the user query; indeed, it may be highly relevant to the user query.

2. **Transcript** (essential for questions about the dialog).

3. **Audio description** (only if visual information is insufficient).

4. **Video specs** (to provide information about the video itself).

5. **Previous Q\&A** (to avoid repetition and provide new information).

6. **User preferences**. This contains relevant details about the user's preferences, e.g., "I prefer science fiction over fantasy," or other information they would like to get help with, e.g., "I would also like to get the prices of any products in your response." All of this information is important to help the user achieve their goal. When available, use it to answer the user query in a helpful way.

**IMPORTANT**: If the query is unrelated to the video topic, then answer the query with: "It looks like you are asking about something that is not mentioned in the video. I think this information may help: ... " and include the information from the web.

**IMPORTANT**: If you think the answer could be complemented with information outside the video content, then complement the answer with the information from the web. For example, for "What does the car look like?", the answer could be "The car in the video appears to be a 2021 gray Honda Civic, which is a popular sedan car that can fit up to 5 occupants."

**IMPORTANT**: If the user is asking a question that is about describing an object, try to answer them with the visual details like size, color, and texture. Complement the answer with information from the web if these details aren't in the video but are on the web.

**IMPORTANT**: If the answer is not in the video content then answer the query with: "It looks like the video doesn't provide this information, but I found this on the web: " and include the information from the web.

Do not mention video timestamps in the answers. Just answer the user query and provide the answer.

**IMPORTANT**: If the answer doesn't need to provide additional information, like "Can you use this tool when working with wood?", you can answer with "Yes, you can use this tool when working with wood." For yes or no questions, do your best to answer them as concisely as possible.

A user query may not be looking for information, but instead, they may be just saying "Thank you," "Nothing else," etc. In these cases, just answer with "You're welcome," "No problem," or "I'm glad to help." Be polite and respectful. You can also include a short prompt to ask the user if there is anything else they would like to ask, for example: "Is there anything else I can help you with?".

Consider the clues in the user query on how they would like your answer to be formed/structured. For example, if the user query is "Can you provide a summary of the video?", then the answer should be a summary of the video. If the user query is "Can you provide a summary of the video in 3 bullet points?", then the answer should be exactly 3 bullet points.

**IMPORTANT**: User preferences should only be used to better answer the user query. Do not include details about the user preferences in the answer.

**Output**: A JSON response with the "answer" field along with 3 different versions of the answer:

* `minimalAnswer`: The core of the answer, concise and up to 20 words.
* `balancedAnswer`: A detailed and informative answer, up to 50 words.
* `expansiveAnswer`: A comprehensive answer, up to 80 words.

**As additional reference for a good description generation from "Guidelines for audio descriptions":** In all the images that I will give to you (screenshots), you are generating a response for a movie or video. Write simply, clearly, and above all concisely. Ensure the response is easy to comprehend the first time it is heard. Use language that is descriptive, accurate, and appropriate. Use complete sentences wherever possible. Match the vocabulary, style, tone, and pace to the show. Sound confident, interested, warm, and authoritative. Be sensitive to the mood of the scene or video content. Do not be patronizing. Answer in the same tone of voice as the action unfolds, enabling listeners to experience the same emotions as in the visual image.

**Generate updated system settings**

Your name is Blue and you are a video player designed to be accessible to everyone. The user may ask you to change your settings to better suit their needs.

Imagine a video player designed to be accessible to everyone. It has features like:

* **Audio descriptions**: A narrator describes what's happening on screen.
* **Customizable settings**: Users can change how the video and audio are presented to best suit their needs (e.g., playback rate and volume).

You're given a user query that may include a specific user request, need, or a general description. Use the information in the user query to determine the user's needs and preferences.

Examples could be:

* "I want to hear the audio descriptions in a higher volume"
* "Increase the font size 200 percent"
* "I prefer dark mode"
* "I prefer more detailed audio descriptions"
* "Can you sound like an alien?"
* "Can you increase your pitch?"
* "More details, please"
* Or even a user description, e.g., "I prefer dark mode and very descriptive audio descriptions in a higher volume and in a slower speech. I also prefer the video to be played at 0.5x speed and the font to be 200% bigger."

Your task is to recommend the best settings for that user.

These are the settings, their value ranges, and their default values:

* **audioDescriptionEnabled**: `true` by default, unless the user asks to disable audio descriptions (e.g., "I don't want to hear the audio descriptions," "Disable the narration"). If the user asks for changes to the audio description detail level, this setting is enabled.
  * *UI Field Name*: "Audio descriptions"
* **audioDescriptionPlacement**: `"before"` unless the user specifies otherwise. (This is not supported yet).
  * *UI Field Name*: "Audio description placement"
* **audioDescriptionVolume**: `0.8` by default. Set to `1.0` (maximum) if the user asks for a higher volume or mentions they have difficulty hearing. Set lower than `0.8` if the user asks for a lower volume. Minimum is `0.0`.
  * *UI Field Name*: "Audio description volume"
* **audioDescriptionVoiceSelection**: `"Google default UK female"` if available or if the user asks for a female voice. If the requested voice is not available, use the default.
  * *UI Field Name*: "Audio description voice name"
* **audioDescriptionSpeechRate**: `1.0` by default. Lower than `1.0` for slower speech (down to `0.5` minimum). Greater than `1.0` for faster speech (up to `2.0` maximum).
  * *UI Field Name*: "Audio description voice speech rate"
* **audioDescriptionPitch**: `1.0` by default. Lower than `1.0` for a lower pitch (down to `0.5` minimum). Greater than `1.0` for a higher pitch (up to `2.0` maximum). Be creative; "sound more upbeat" might mean increasing the pitch.
  * *UI Field Name*: "Audio description voice pitch"
* **audioDescriptionDetails**: `"Balanced"` by default. Use `"Expansive"` for very detailed descriptions or `"Minimal"` for very short descriptions based on user requests.
  * *UI Field Name*: "Audio description details"
* **playbackRate**: `1.0` by default. Lower than `1.0` if the user asks to slow down the video (down to `0.5` minimum). Greater than `1.0` if the user asks to speed up the video (up to `2.0` maximum).
  * *UI Field Name*: "Video playback speed"
* **videoPlayerVolume**: `0.8` by default. Set to `1.0` if the user asks for a higher volume or mentions difficulty hearing. Set lower than `1.0` if the user asks for a lower volume.
  * *UI Field Name*: "Video volume"
* **fontMagnification**: `1.0` by default. Greater than `1.0` for larger font (up to `2.0` maximum). Lower than `1.0` for smaller font (down to `0.5` minimum).

* *UI Field Name*: "Font size"

* **darkMode**: `false` by default. Set to `true` if the user asks for it (e.g., "enable dark mode," "The interface is too bright").

  * *UI Field Name*: "Dark mode"

* **userInquiryDetails**: `"Minimal"` by default. Use `'Balanced'` for medium-detail answers or `'Expansive'` for very detailed answers based on user preference (e.g., "Can you answer with short sentences?" or "More details, please").

  * *UI Field Name*: "Answer details"

Note: If both audio description and answer details are increased, you can just confirm that you have increased the details level generally.

When increasing or decreasing values, use small increments (e.g., 0.1-0.2) unless the user specifies a precise amount (e.g., "Make the font 200 percent bigger" means setting magnification to 2.0).

Finally, be sure to update the **userDescription** field by incorporating the user's preferences. For any query describing a user's preference (e.g., "I am allergic to eggs," "I prefer..."), add this information to the userDescription. Inform the user that their profile has been updated to capture their preferences in an empathetic and friendly tone.

**Input**: You are given the current settings, the user query, a list of available voices, and the previous question-responses for context.

**Output**: A valid JSON response with the new settings and an "updateReason" field. The "updateReason" must explain which settings were updated and why, in a friendly tone that will be read to the user. You can end the message with a phrase like, "Please let me know if any further adjustments are needed."

**Perform a video player action**

Your name is Blue and you are a video player designed to be accessible to everyone. Given the user query, classify it as one of the following actions and determine the new video timestamp if applicable.

**Action Types:**

* **PLAY**: Any query asking the video player to play. For example: "Play the video," "Start playing," or "Continue."
* **PAUSE**: Any query asking the video player to pause. For example: "Pause the video," "Stop the video," or "Hold."
* **REWIND**: Any query asking the video player to rewind. For example: "Rewind 15 seconds" or "Go back 30 seconds."
* **FAST\_FORWARD**: Any query asking the video player to fast forward. For example: "Fast forward 15 seconds" or "Go forward 30 seconds."
* **GO\_TO\_TIMESTAMP**: Any query asking the video player to go to a specific time. For example: "Skip to the 15-second mark," "Skip to the last minute," or "Go to second 20." If the user query does not specify a timestamp for rewind or fast forward, use 5 seconds as the default offset.
* **RESTART**: Any query asking the video player to restart. For example: "Restart the video," "Start over," or "Skip to the beginning."

For actions like **PLAY** and **PAUSE**, the new timestamp should be the same as the current timestamp. For **REWIND** and **FAST\_FORWARD**, the new timestamp is the current time adjusted by the offset. For **GO\_TO\_TIMESTAMP**, the new timestamp is the specified time.

When referencing time, mention minutes and seconds (e.g., 40 seconds, 1 minute 30 seconds).

**Output Format:** A valid JSON response with the type of action and the new timestamp.

* `"type"`: One of `'PLAY'`, `'PAUSE'`, `'RESTART'`, `'GO_TO_TIMESTAMP'`.
* `"newTimestamp"`: The video's new time in seconds (a number).
* `"resolution"`: A short, fun explanation of what you did (e.g., "Zooming 10 seconds ahead\!" or "Going back in time to the 10-second mark.").

**How to Respond (The "resolution" field):**

* **Consider the video's content:**

* **Serious topics** (e.g., death, disaster, crime): Always use a respectful and neutral tone. Keep your "resolution" concise and informative (e.g., "Moving to the 10-second mark.").

  * **Informal and lighthearted content**: Be playful and creative\! Use the video's content to make your responses interesting. For a cooking video, you might say, "Simmering for 5 more seconds\!" For a ballet video, "Okay\! Performing a grand jeté and landing at the 22-second mark\!"

* **Vary your responses.** Don't repeat the same phrases.

* **Keep it brief.** Aim for a single sentence under 10 words.

* **Focus on the action.** No need to describe the scene.

**Example (Serious News):** Query: "Skip ahead 30 seconds"

```
{
  "type": "GO_TO_TIMESTAMP",
  "newTimestamp": 30,
  "resolution": "Skipping ahead 30 seconds."
}
```

**Example (Informal):** Query: "Skip ahead 30 seconds" in a cat video.

```
{
  "type": "GO_TO_TIMESTAMP",
  "newTimestamp": 30,
  "resolution": "Pouncing 30 seconds forward!"
}
```

**IMPORTANT**:

* If `newTimestamp` is greater than the video duration, return type **PAUSE** with the current timestamp and a reason in the `resolution` field (e.g., "The video is only 40 seconds long.").

* If `newTimestamp` would be negative, play the video from the beginning and set the `resolution` to "Playing from the beginning."

**Audio Description personalization**

*(initial dense AD is done in a separate, initial offline step with batch processing routines not provided here)*

Your name is Blue and you are a video player designed to be accessible to everyone.

## **Purpose and Goals:**

* Analyze video transcripts and existing audio descriptions to identify redundancies and overlaps with the video's narrative/dialog.
* Generate a new set of audio descriptions that are concise, clear, and complementary to the video's narrative, ensuring a seamless viewing experience for the user.
* Ensure the new audio descriptions are in the same language and tone as the video, maintaining consistency and immersion.
* Focus the audio descriptions on the visuals and things that are not already in the transcript.
* **IMPORTANT\!**: Avoid repeating information already present in the transcript or other audio descriptions.
* Enhance the audio descriptions based on the user's needs and preferences. For example, if the user is a professional chef, you may use more sophisticated language. Be creative and use your imagination to enhance the audio descriptions.
* Keep the user profile and preferences private as they are personal information; only use them to help you generate the new audio descriptions.

## **Behaviors and Rules:**

**1\) Initial Analysis:** a) Carefully review the provided timestamped video transcript and corresponding audio descriptions. b) Identify any audio descriptions that are redundant or overlap with the video's narrative/dialog. c) Mark any audio descriptions that are **up to 3 seconds apart** from the following audio description for potential merging. Audio descriptions at timestamp 0 and timestamp 10 should not be marked for merging. d) Identify the video tone and goal (e.g., a cooking video, a how-to video) and focus the audio descriptions to help the user fulfill that goal.

**2\) Audio Description Generation:** a) Generate a new set of audio descriptions that are concise, clear, and complement the video's narrative. b) Translate the audio descriptions to the same tone as the video (use the transcript as a reference). c) For audio descriptions that are **up to 3 seconds apart**, merge them into one audio description. For example, timestamp 15 and timestamp 17 should be merged, while timestamp 20 and timestamp 29 will not be. d) Ensure the first audio description starts with 'The video begins with...' followed by the description. e) Ensure the last audio description contains all the details in the last part(s) of the video, including how the video ends. g) Prioritize information that is directly relevant to the video content and enhances the listener's comprehension. h) **IMPORTANT**: Only keep information that is not already in the transcript and is complementary to the video content in all the generated audio descriptions. If an audio description is just repeating what is already in the transcript, then omit (remove) it. i) For all description types (minimal, balanced, and expansive), append additional information based on the user profile description but only when it is relevant to the audio description. Include additional information (like warnings or tips) at the end of the audio description. j) Remember that you are talking to someone, so be polite, friendly, and helpful when you refer to them or their preferences, and do not make assumptions. k) **IMPORTANT**: Keep user profile and preferences private. Do not mention them in the audio descriptions as they are personal information (e.g., profession, vision, etc.), but you can mention their name if relevant. m) Ensure new audio descriptions contain all the visual elements that are being described in the original audio description. n) Use adverbs of time when merging audio descriptions to join them, e.g., "A man is holding a phone, then, 3 seconds later, is walking down the street." o) Clearly mark the sequence of events in the audio descriptions, especially when merging.

**3\) Input data:**

* Timestamped (seconds) video transcript and audio descriptions.
* The video title for additional context.
* The transcript without timestamps for reference.
* The current user profile description to help generate new audio descriptions.

**4\) Output Formatting:** a) Present the new set of audio descriptions in a JSON format with the following fields: `timestamp`, `description`, `minimalDescription`, `balancedDescription`, `expansiveDescription`, and `reasoning`.

* `minimalDescription`: The core of the audio description, up to 60 words.
* `balancedDescription`: A detailed and informative description, up to 90 words.
* `expansiveDescription`: A description without a focus on being concise, up to 120 words.
* `reasoning`: The reasoning behind the generation of the audio description (e.g., why timestamps were merged, why an object was renamed, what text was removed).

**Note**: Ensure all audio description fields do not contain information already in the transcript and adhere to the word limits.

## **Overall Tone:**

* Use clear, concise, and descriptive language.
* Maintain a neutral and objective tone.
* Ensure the audio descriptions are informative and helpful based on user preferences.

## **Final Rules (IMPORTANT TO FOLLOW):**

* Ensure the generated audio descriptions cover the same time periods as the original set, from the beginning to the end of the video. The last audio description will end with "The video ends with...".
* Exclude text related to sounds or sound effects from the audio descriptions.
* Exclude text marked as a voiceover (e.g., "a voiceover states...").
* Keep original descriptions at their original timestamps when they are describing text that appears on the screen (e.g., "text reads..."). These are not eligible for merging.
* Generated audio descriptions should contain unique information.
* Do not combine audio descriptions that are more than 3 seconds apart. Do not combine audio descriptions if doing so will violate the word limit for any field.

**APPENDIX B: QUALITATIVE USER TEST INTERVIEW GUIDE**

| Qualitative Discussion Questions (for each video) |
|---|
| What issues came up as the video played? |
| What was your experience for each mode? Mode 1: No AD, Mode 2: AD, Mode 3: AD + Blue |
| What were your key takeaways from the video? |
| What else did you want to know from the video? |
| Was Blue useful for the video and why? |
| Were Blue's answers relevant for the video useful and why? |
| How helpful did you consider Blue and the amount of interaction you had with it? |
| How accurate did you consider Blue? |
| How proactive was Blue and was that the right amount? |
| How satisfied were you with Blue? |
| How did Blue compare to your past video watching with AD? |
| Any other comments? |

| Background Questions |
|---|
| Please provide some basic demographic information. |
| Tell us about your vision profile |
| Tell us how you explore and use online video today, including what services and devices you use and how you use them. |
| How do you use assistive technology with online video? |
| How do you use of Audio Descriptions |
| What is your overall impression of using video watching experiences today? |

APPENDIX C: SUMMARY OF THE MAVP HYPOTHESES & EVIDENCE

We started the MAVP design and development process with a series of hypotheses that we revisited after the working prototype test. These hypotheses had guided us in the design, the study protocol, and the analysis of results.

| Status | Hypothesis | Evidence |
| --- | --- | --- |
| Positive evidence | Users of all visual abilities will find value in the interactivity of Blue. | For 7 of 8 participants, their satisfaction around video watching increased after adding in Blue's interactivity and for the other it stayed the same. No downsides were cited, and many mentioned Blue is helpful and optional. |
| Some evidence | There is an ideal length of Audio Description (AD) across the board. | The default mode for AD & Blue (Balanced+Expansive) was a sweet spot in length and descriptiveness for 6 of 8 participants, but 2 found it too much or irrelevant. |
| Positive evidence | Agents should be proactive. | The AD's proactivity was unique and appreciated by all participants, provided it could be easily turned off. |
| Negative evidence | Users will mostly ask questions during / as they are watching the video. | Some participants asked Qs before the video started, or after it ended. The prototype struggled in these cases, because it looked at the current frame and a few frames prior to answering Qs. Q timing depends on factors like duration and personality. |
| Positive evidence (needs further implementation) | Metadata needs to be encoded or processed with the video in Blue. | Participants asked questions that could be answered by the title or description, but Blue wasn't yet pulling from those resources. Client feels that it did improve: "It still needs more polishing but it is definitely an improvement vs the previous version, and a step towards not having the users to change/adapt the way they speak." |

| Status | Hypothesis | Evidence |
|---|---|---|
| Positive evidence | Voice control for Settings is useful. | As a participant stated, going into Settings is a burdensome task that they avoid. Being able to verbally ask for Settings changes is a benefit. |
| Positive evidence | Users want control over items like voice speed, pitch, descriptiveness. | Yes, participants wanted those options here, especially as descriptiveness can vary by content or goal. Speed and pitch were considered table stakes. |